\documentclass[a4paper,fleqn,usenatbib]{mnras}

\usepackage{newtxtext,newtxmath}

\usepackage[T1]{fontenc}
\usepackage{ae,aecompl}

\usepackage{graphicx}	
\usepackage{amsmath}	
\usepackage{amssymb}	

\title[The distances to $\chi$~Per, NGC~7419 and Wd~1]{The distances to star-clusters hosting Red Supergiants: \\$\chi$~Per, NGC~7419, and Westerlund 1}

\author[Davies and Beasor]{
Ben Davies$^{1}$\thanks{b.davies@ljmu.ac.uk} and Emma R. Beasor$^{1}$
\\
$^{1}$Astrophysics Research Institute, Liverpool John Moores 
University, Liverpool Science Park ic2, 146 Brownlow Hill, Liverpool, L3 5RF, UK\\
}

\date{Accepted XXX. Received YYY; in original form ZZZ}

\pubyear{2019}

\begin{document}
\label{firstpage}
\pagerange{\pageref{firstpage}--\pageref{lastpage}}
\maketitle

\def\dwdone{\hbox{$d=3.87^{+0.95}_{-0.64}$\,kpc}}
\def\dchiper{\hbox{$d=2.25^{+0.16}_{-0.14}$\,kpc}}
\def\dngc{\hbox{$d=3.00^{+0.35}_{-0.29}$\,kpc}}
\def\pingc{\hbox{$\bar{\pi} = 0.334 \pm 0.018$\,mas}}
\def\piwdone{\hbox{$\bar{\pi} = 0.259 \pm 0.036$}\,mas}
\def\pichiper{\hbox{$\bar{\pi} = 0.444 \pm 0.006$}\,mas}

\begin{abstract}
Galactic, young massive star clusters are approximately coeval aggregates of stars, close enough to resolve the individual stars, massive enough to have produced large numbers of massive stars, and young enough for these stars to be in a pre-supernova state. As such these objects represent powerful natural laboratories in which to study the evolution of massive stars. To be used in this way, it is crucial that accurate and precise distances are known, since this affects both the inferred luminosities of the cluster members and the age estimate for the cluster itself. Here we present distance estimates for three star clusters rich in Red Supergiants ($\chi$~Per, NGC~7419 and Westerlund~1) based on their average astrometric parallaxes $\bar{\pi}$ in Gaia Data Release 2, where the measurement of $\bar{\pi}$ is obtained from a proper-motion screened sample of spectroscopically-confirmed cluster members. We determine distances of \dchiper, \dngc, and \dwdone\ for the three clusters respectively. We find that the dominant source of error is that in Gaia's zero-point parallax offset $\pi_{\rm ZP}$, and we argue that more precise distances cannot be determined without an improved characterization of this quantity. 
\end{abstract}


\newcommand{\fig}[1]{Fig.\ \ref{#1}}
\newcommand{\Fig}[1]{Figure \ref{#1}}
\newcommand{\newtext}[1]{{\color{blue} #1}}


\section{Introduction}
Historically, star clusters have been used as natural laboratories in which to test the theory of stellar evolution. This is particularly true for massive stars, where the short lifetimes and non-monotonic mass-luminosity relation make it very difficult to infer the evolutionary state of isolated stars. In clusters where the age and distance are known, it is possible to constrain the initial masses of a wide variety of post main-sequence objects. For example, it has been possible to constrain the nature of the Of/WNh stars \citep{Martins08}, infer the progenitor masses of the progenitors of neutron stars \citep{sgr1900paper},  argue for high mass progenitors to Wolf-Rayet (WR) stars (provided membership with and age of the host cluster can be firmly established) \citep[e.g.][]{Humphreys85,Massey01,Clark05}, and measure an accurate mass-loss rate law for Red Supergiants (RSGs) \citep{Beasor-Davies16,Beasor-Davies18}. 

All of the above studies rely on being able to obtain accurate ages, reddenings and distances to the host star clusters. The latter quantity (and reddening, in the case of high foreground extinction) is vital in determining the bolometric luminosities of the cluster stars, allowing them to be placed on a diagnostic H-R diagram. Once the luminosities of the turn-off and post-MS stars are known, an age may then be inferred, although this process itself has many pitfalls \citep{Beasor2019}. Accurate distances to young star clusters are therefore pivotal to understanding the evolution of massive stars. 

Distances to young massive star clusters in the Milky Way have typically been estimated via three independent methods. If the cluster radial velocity can be measured, for example from the average of the member stars or from the surrounding interstellar medium, a kinematic distance may be inferred by comparing to the Galactic rotation curve \citep[e.g.][]{rsgc1paper,Kothes-Dougherty07}. If the cluster has a low foreground extinction, deep optical imaging can reveal the `kink' in the main-sequence caused by the transition from the PPI-chain to the CNO-cycle as the main form of energy generation, which can be used as a distance-sensitive anchor for isochrone fitting \citep[e.g.][]{Currie10}. Finally, if spectroscopic observations can go deep enough to detect the more well-behaved main sequence stars of spectral type late-O / early-B, spectroscopic parallaxes may be obtained \citep[e.g.][]{danks-paper,Crowther06}.

Until recently, the much more direct method of obtaining distances, from their astrometric parallaxes, was not possible for Galactic YMCs. Such objects are relatively rare, and so typically have distances $>$2kpc, requiring parallax measurements accurate to better than 0.1mas. Furthermore, at these distances there is often substantial reddening, compounding the problem. The second data release of Gaia (DR2) \citep{Gaia,GaiaDR2} therefore represents an opportunity to revolutionise the field of massive star research, as distances to several benchmark clusters and associations may now be obtained at much higher accuracy and precision than has previously been possible. 

In this paper, we focus on three star clusters young and massive enough to contain several Red Supergiants, and whose cluster members are bright enough in the optical and sufficiently uncrowded to have reliable detections in DR2. These clusters are $\chi$~Per, NGC~7419, and Westerlund~1. In Sect.\ \ref{sec:method} we describe our methodology in terms of how we select the benchmark cluster members and determine an average cluster parallax. In Sect.\ \ref{sec:results} we present the results, and conclude in Sect.\ \ref{sec:conc}.

\section{Method} \label{sec:method}
\subsection{Sample definition}
We begin by searching the SIMBAD\footnote{} database for OB stars within 0.5\degr\ of the centre of each cluster. We concentrate on OB stars, as the parallax measurements of late-type supergiants are known to be problematic owing to the size of the stars being comparable to (or greater than) the size of the Earth's orbit around the Sun \citep[see e.g.][]{Chiavassa11gaia}. We then cross-match this sample with Gaia~DR2, to obtain parallaxes $\pi$, proper motions (PMs), and associated errors. Following \citet{Lindegren18} and \citet[][ hereafter A19]{Aghakhanloo19}, we define the error on the parallax $\sigma_i$ of each star $i$ to be $\sigma_i = 1.086\sigma_\pi$ where $\sigma_\pi$ is the quoted error on $\pi$ in Gaia DR2. 


Next, we isolate those stars with proper motions (PMs) consistent with the cluster average. This allows us to eliminate stars with potentially anomalous parallaxes, such as runaways or binaries. We define the average PM for each cluster by performing an iterative sigma-clipped mean using the IDL function {\tt meanclip}, clipping at 1.5$\sigma$. We then isolate those stars within 2.5$\sigma_{\rm PM,i}$ of this mean, where $\sigma_{\rm PM,i}$ is the error on each star's PM. We deliberately set these tight (and potentially exclusive) constraints as we are not concerned with being complete, only with identifying the stars with reliable astrometric information. The results of this process are illustrated in \fig{fig:PMpi}. We define the remaining stars as the `clean' samples, which contain 62, 10 and 32 stars for the clusters $\chi$~Per, NGC~7419, and Wd~1 respectively. The sensitivity of our results to how aggressively we perform the PM cleaning are discussed in Sect.\ \ref{sec:results}.

\subsection{Average cluster parallax, $\bar{\pi}$} \label{sec:pi}
\begin{figure*}
\begin{center}
\includegraphics[width=5.8cm]{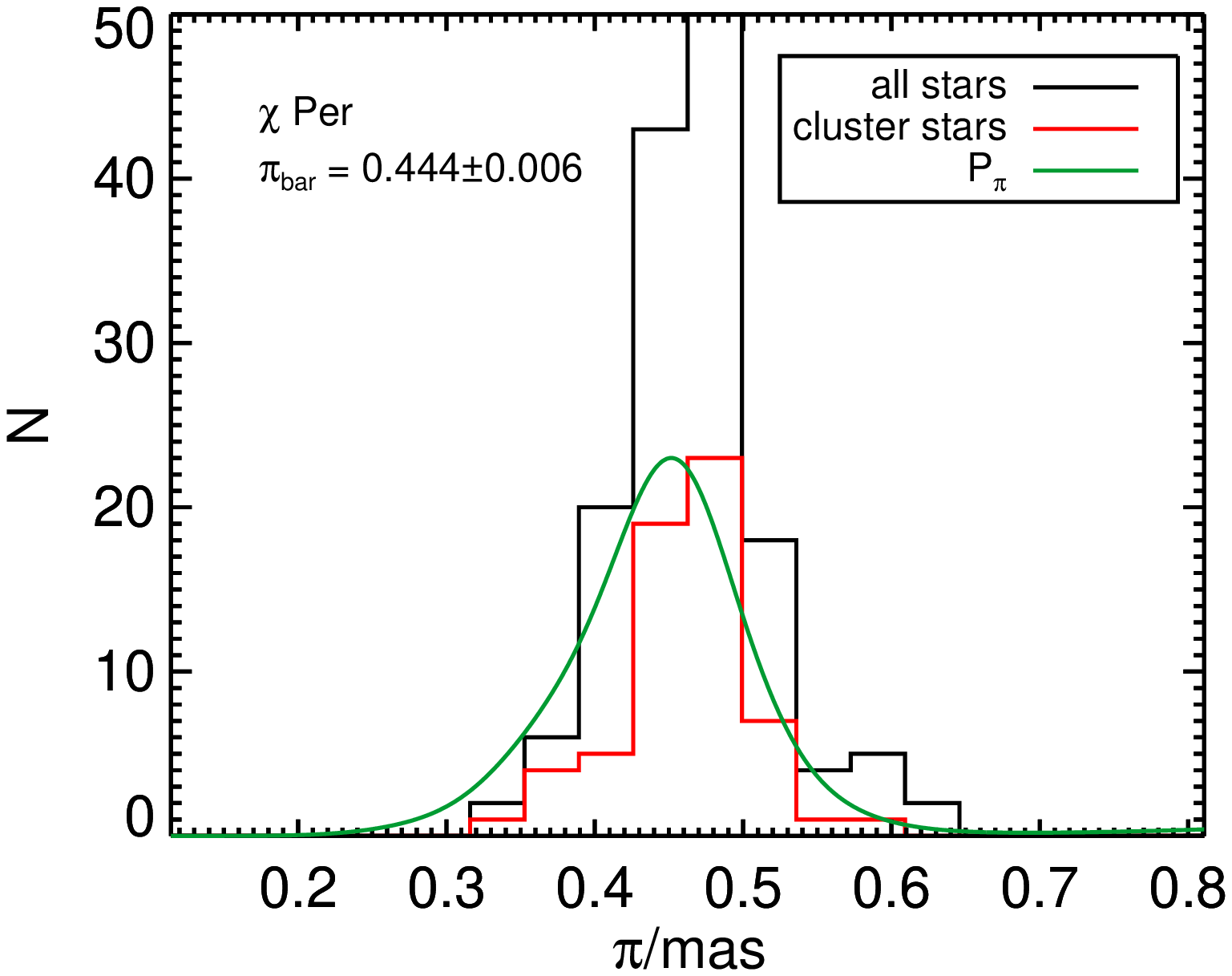}
\includegraphics[width=5.8cm]{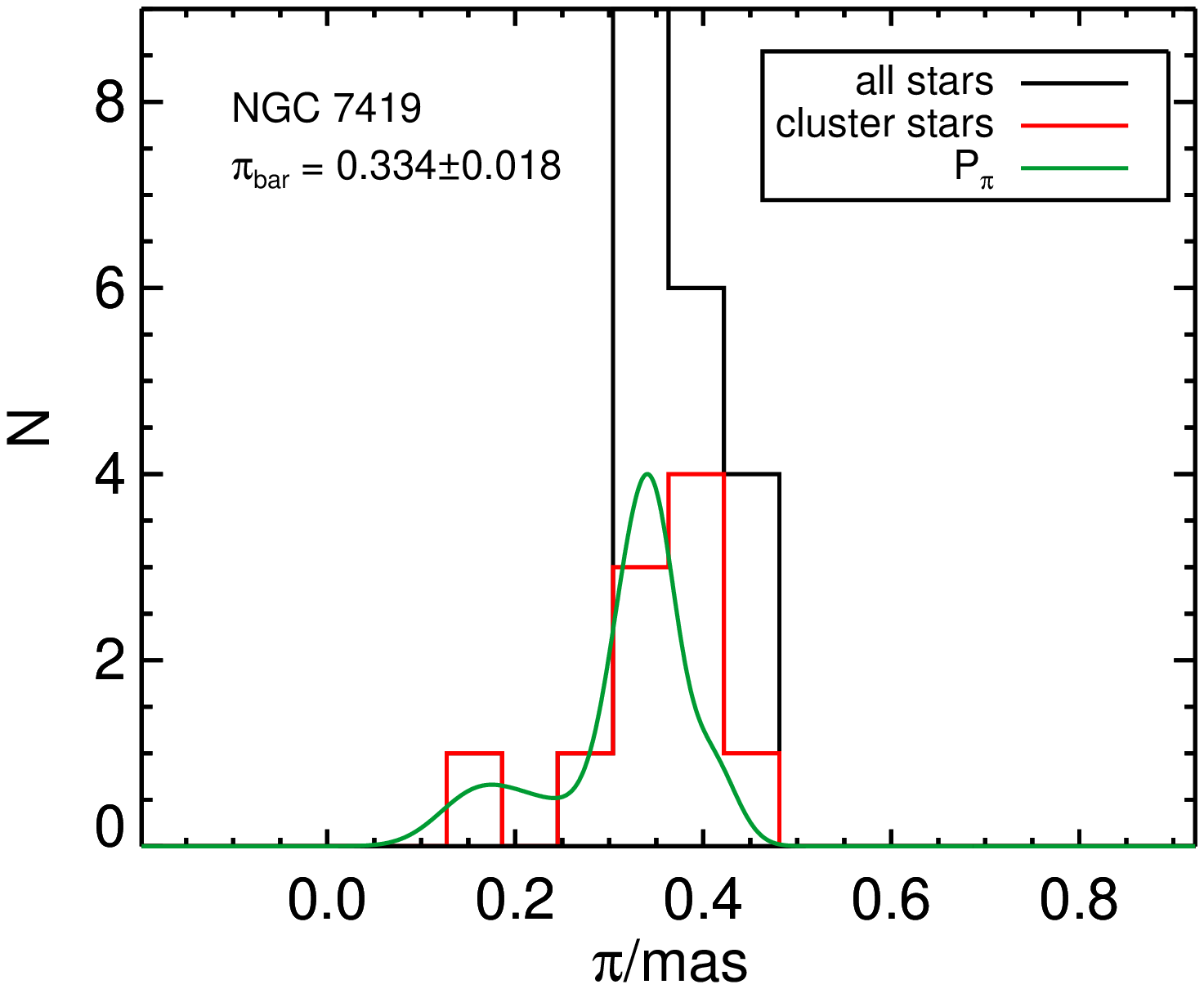}
\includegraphics[width=5.8cm]{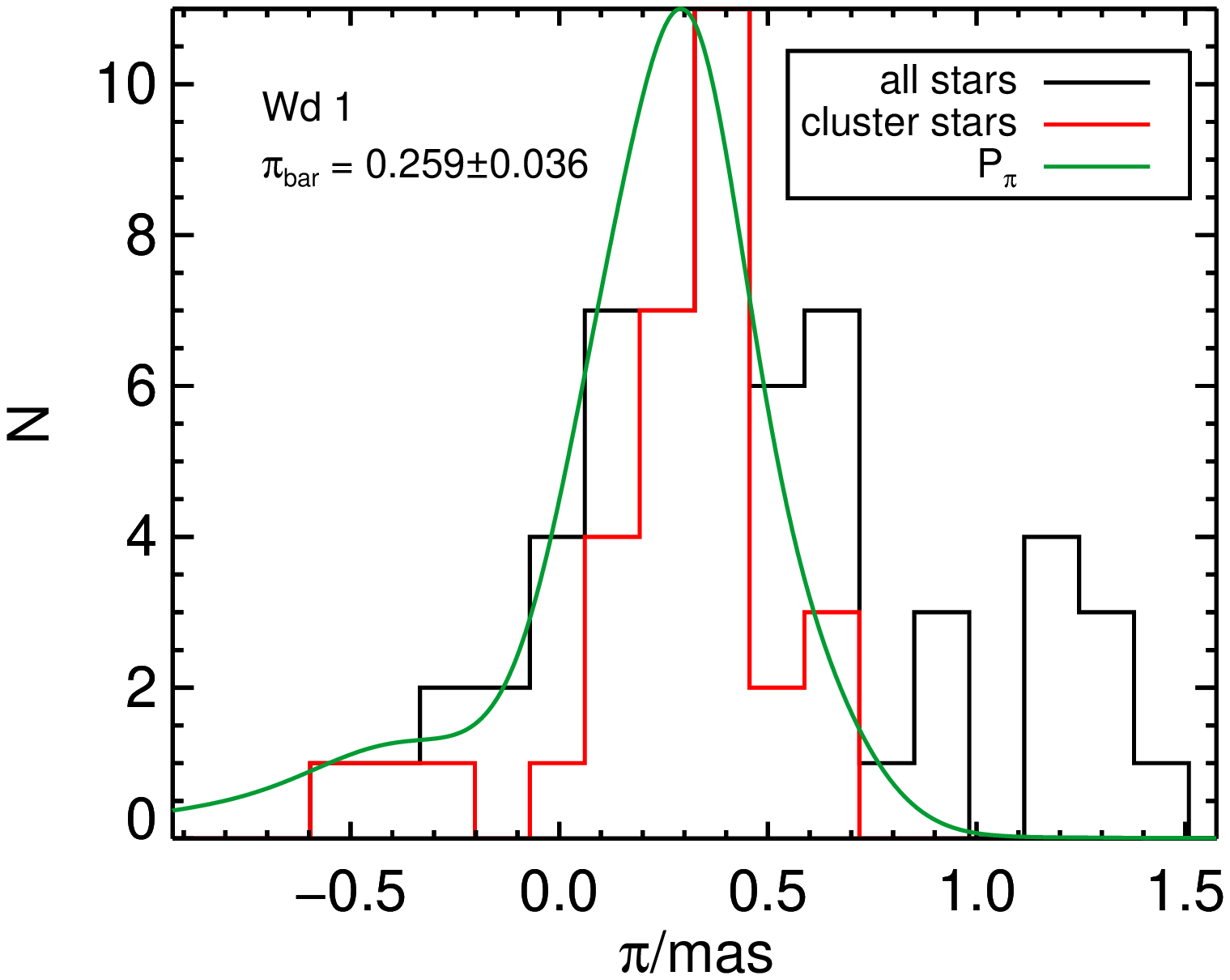}
\caption{Parallaxes of the OB stars in the field of each cluster. Black lines show the histograms of all stars in the field; red lines show the same but only for the clean sample; and green shows the total probability distribution for the average parallax, which takes into account the error bars on each parallax measurement. The weighed mean parallax is shown as the green dashed line. The average zero-point parallax offset of -0.05mas has been applied to all stars (see text for details).}
\label{fig:Ppi}
\end{center}
\end{figure*}

The next step is to define the average parallax $\bar{\pi}$ to each cluster. In \fig{fig:Ppi} we plot histograms of the parallaxes of the stars in each cluster field. We plot all OB stars in the fields of the clusters in black, and the cleaned sample in red. In each case, the parallaxes are somewhat normally distributed. This is just as one would expect if each star's parallax were randomly sampled from a gaussian distribution centred on the mean cluster parallax with a standard deviation characteristic of the error on each measurement. In fact, the errors on each $\pi_i$ are not all the same. To get a more representative illustration of the distribution of parallaxes, we determined the probability distribution functions for each $\pi_i$, assuming a gaussian distribution with width $\sigma_i$, and summed over all stars to determine the total $\pi$ probability function, $P_\pi$ . The results are shown in the green curves of \fig{fig:Ppi}. The green dashed lines in these figures are the weighted means of the cleaned samples, which we call $\bar{\pi}$. We determine $\bar{\pi}$ and its error $\delta\bar{\pi}$ according to,

\begin{equation}
\bar{\pi} = \frac{\sum^{N}_{i} w_{i} \pi_{i}}
{\sum^{N}_{i} w_{i}} , ~~~
\delta\bar{\pi} = \sqrt{ \frac{1}{N-1}
\frac{ \sum^{N}_{i} w_{i} (\pi_{i} - \bar{\pi})^2 } {\sum^{N}_{i} w_{i}} }
\end{equation}

\noindent where $N$ is the number of stars in the `clean' sample, and the weights $w_i = 1/\sigma_i^2$. Note that the error on the mean $\delta\bar{\pi}$ is the weighted standard deviation divided by $\sqrt{N}$. 

\subsection{Distance, $d$} \label{sec:dist}
To convert $\bar{\pi}$ to a distance $d$, we first determine the posterior probability distribution on $\bar{\pi}$, 

\begin{equation}
\displaystyle
P_d \propto \exp \left({-\frac{1}{2} z^2}\right)
\end{equation}

\noindent where,

\begin{equation}
z = \frac{ \bar{\pi} - \pi_{\rm ZP} - 1/d}{\delta\bar{\pi}}
\end{equation}

\noindent and $\pi_{\rm ZP}$ is the zero-point parallax offset in Gaia~DR2. 

The quantity $\pi_{\rm ZP}$  has been studied by numerous authors using several independent methods, with values ranging from -0.029mas$<\pi_{\rm ZP}<$-0.08mas \citep{Stassun18,Riess18,Lindegren18,Graczyk19}. Specifically, Lindegren et al.\ found that the this offset varied with position on the sky with an amplitude of $\delta\pi_{\rm ZP}\pm0.03$mas on spatial scales of about a degree. Since this fluctuation occurs on a spatial scale larger than the apparent size of our clusters, it must be assumed to affect all stars equally, That is, the error on $\pi_{\rm ZP}$ fixes a lower limit to the uncertainty on the absolute parallax of the cluster. With this in mind, we add the quantities $\delta\pi_{\rm ZP}$ and $\sigma_{\bar{\pi}}$ in quadrature when determining the absolute uncertainty on $\bar{\pi}$. Throughout this work we adopt an average value $\pi_{\rm ZP} = -0.05\pm0.03$mas.

Having calculated $P_d$, the distance $d$ and uncertainty $\sigma_d$ are determined from the mode and 68\% confidence intervals on $P_d$. Note that, in contrast to other studies which attempt to determine distances from Gaia parallaxes, we do not apply a prior on distance when determining the posterior probability distribution.

\begin{figure*}
\begin{center}
\includegraphics[width=5.8cm]{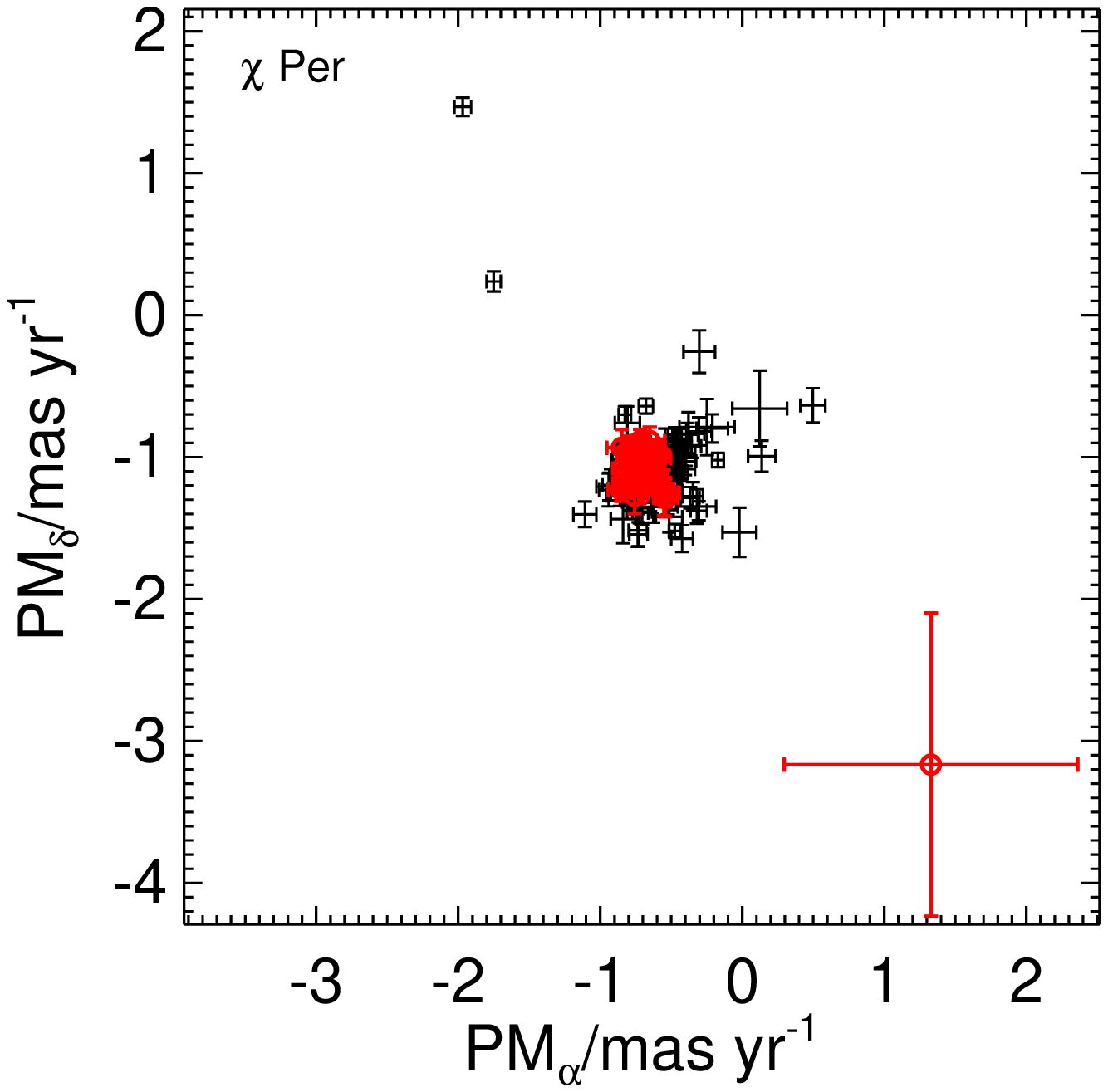}
\includegraphics[width=5.8cm]{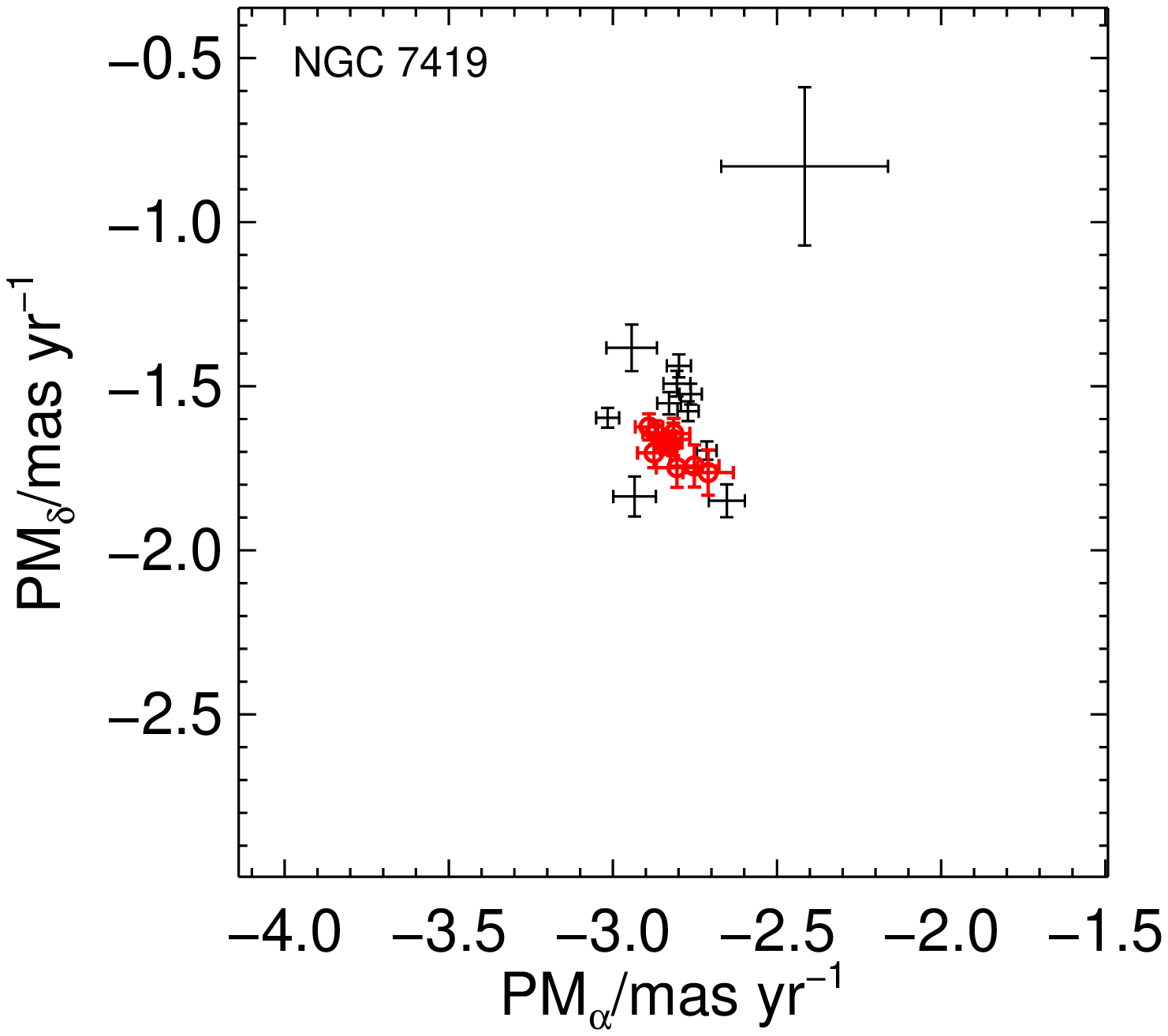}
\includegraphics[width=5.8cm]{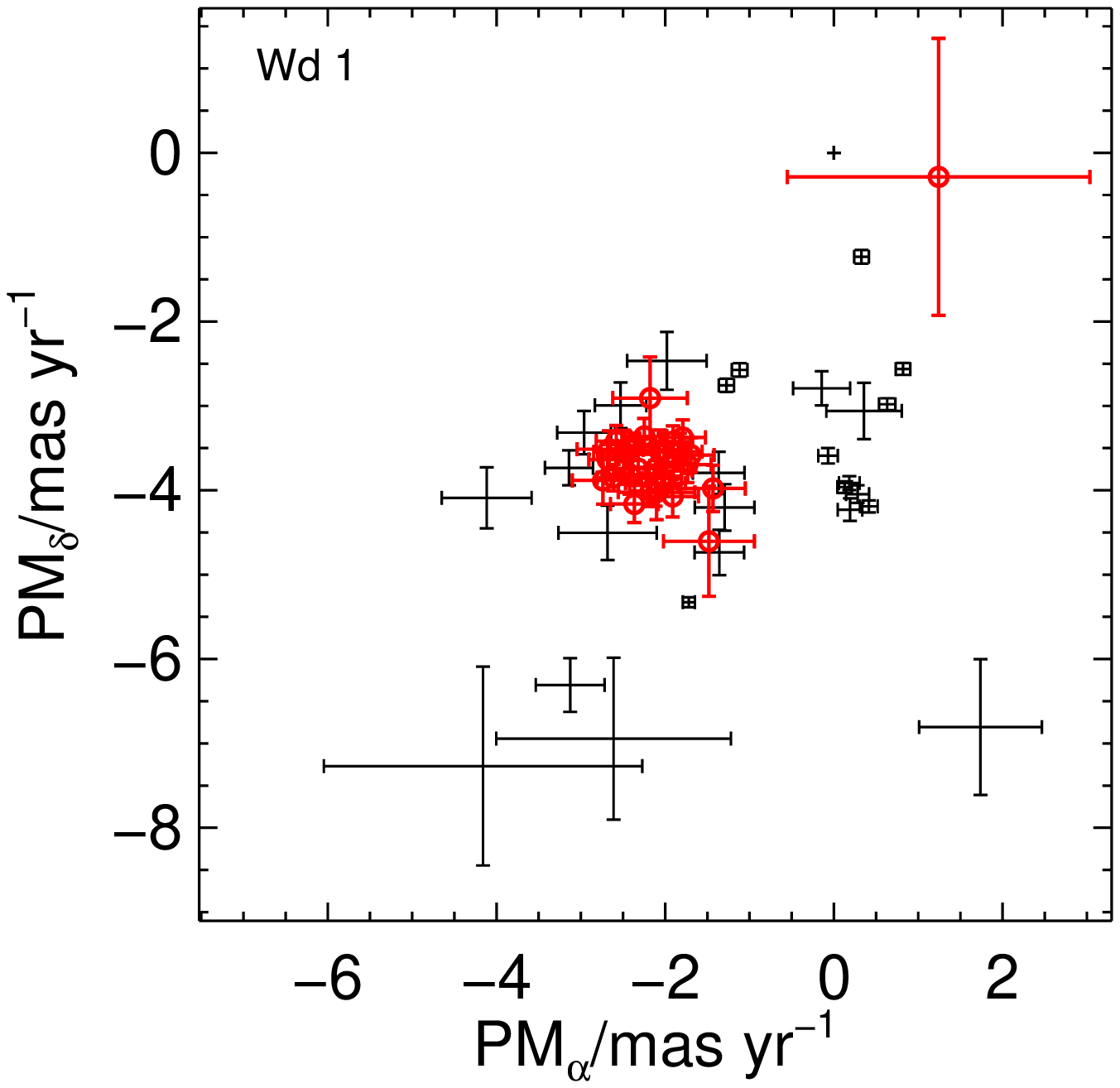}
\caption{Proper motions of the OB stars in the plane of each cluster. Stars deemed to be cluster members with high confidence based on their proper motions (the `clean' sample) are plotted as red circles (see text for details).}
\label{fig:PMpi}
\end{center}
\end{figure*}

\begin{figure*}
\begin{center}
\includegraphics[width=5.8cm]{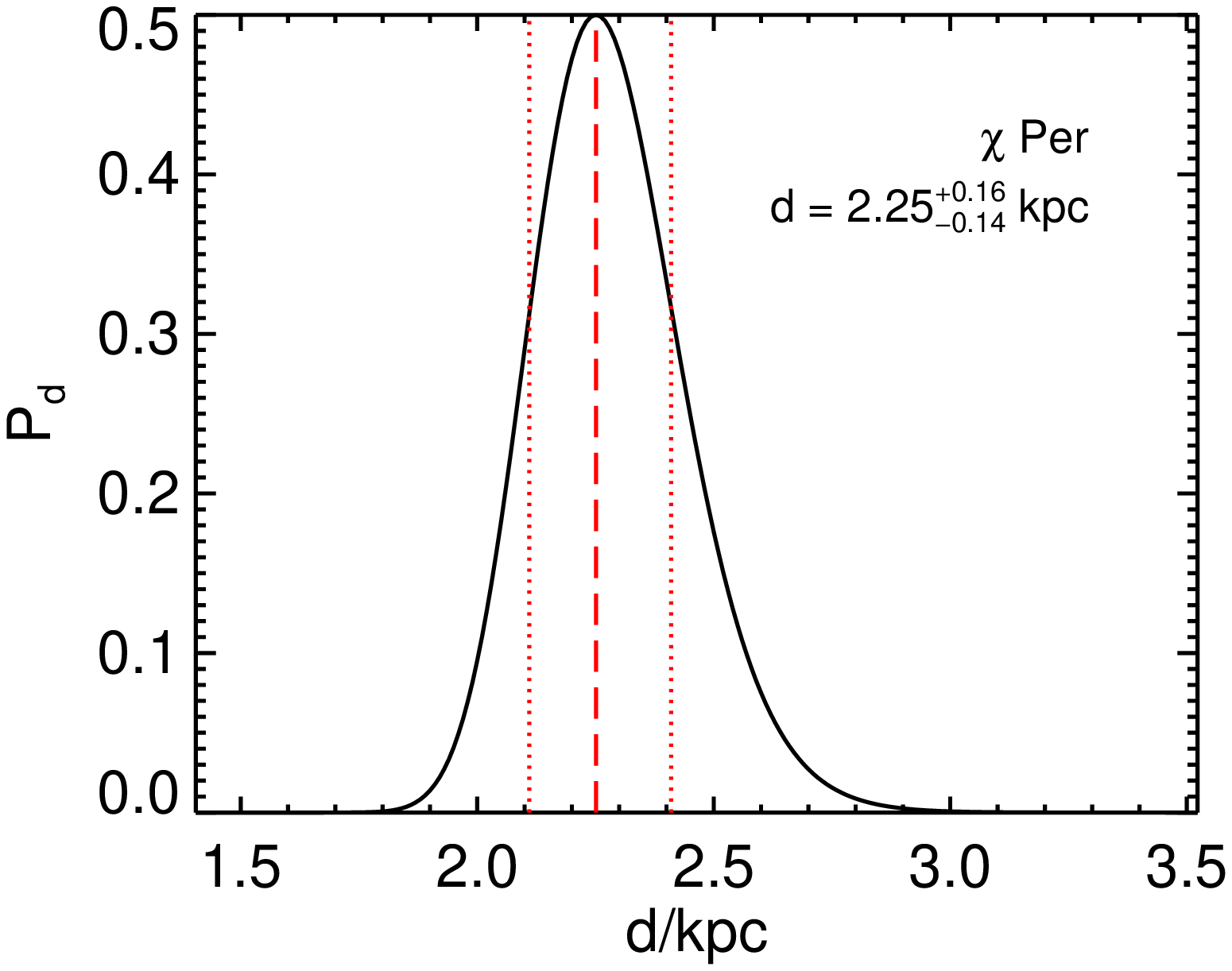}
\includegraphics[width=5.8cm]{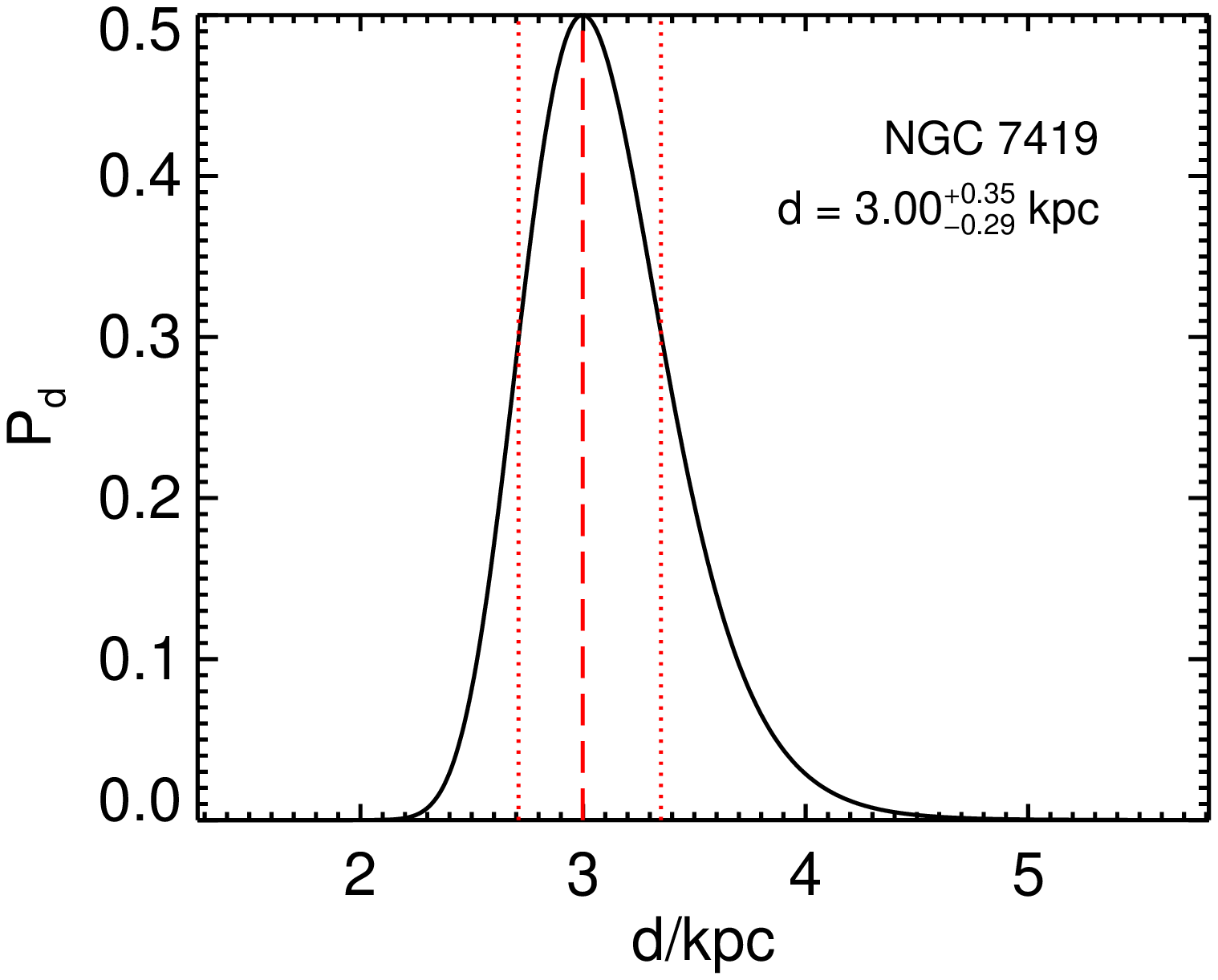}
\includegraphics[width=5.8cm]{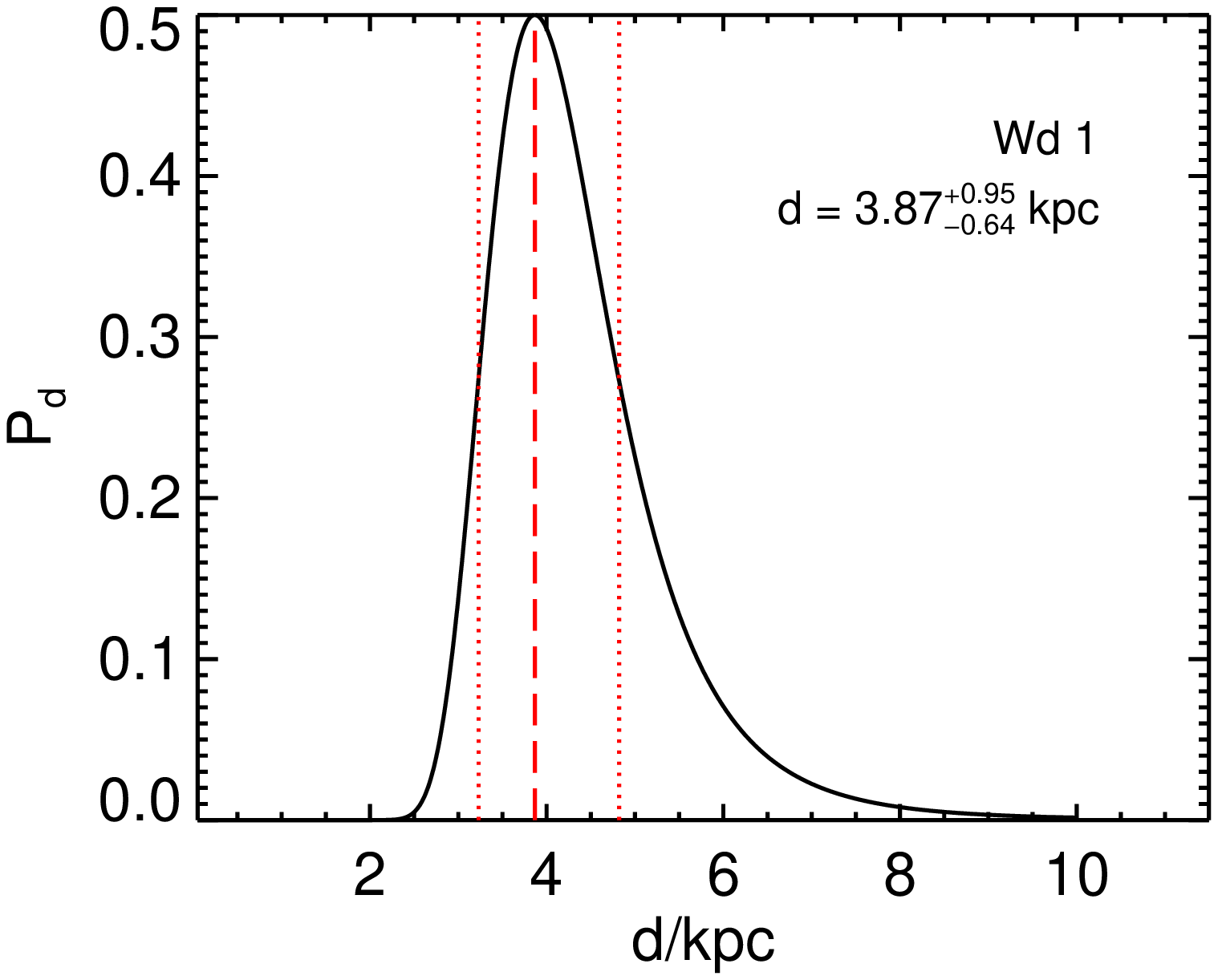}
\caption{Posterior probability distribution of the distances to each cluster. The dashed lines represent the most probable distance and the 68\% confidence intervals.}
\label{fig:dist}
\end{center}
\end{figure*}

\section{Results and discussion} \label{sec:results}
We now discuss our results for the average parallaxes and distances to the three clusters in our sample. The implications for the ages of these clusters, and therefore for how their stellar populations reconcile with stellar evolutionary theory, is complex since it also depends on how one defines the age. This is discussed in a companion paper \citep{Beasor2019}.

\subsection{$\chi$ Persei}
Previous estimates of this cluster's distance have involved fitting the (pre-) main-sequence population in one form or another. The state-of-the-art was presented in \citet{Currie10}. These authors fit the main sequence in a variety of colours and magnitudes, as well as obtaining spectrophotometric distances from the stars with known spectral types, and consistently found a distance within the range $2.344^{+0.088}_{-0.085}$\,kpc. This compares well to similar analysis by \citet{Slesnick02}, \citet{Uribe02}, and \citet{Mayne-Naylor08}.

An alternative estimate to the distance to $\chi$~Per can be found from the maser parallax measurement of the Red Supergiant S~Per. Though unlikely to be a member of the cluster itself (projected distance = 1.47\degr\ $\simeq$60pc at a distance of 2.25kpc), it does belong to the larger Perseus OB1 association, of which $\chi$~Per is also a member. In an astrometric study of the H$_2$O masers around S~Per, \citet{Asaki10} found proper motions of (\hbox{$\alpha$=-0.49$\pm$0.23mas\,yr$^{-1}$}, \hbox{$\delta$=-1.19$\pm$0.20mas\,yr$^{-1}$}), and a parallax of \hbox{$\pi = 0.413\pm0.017$mas}. This is within the errors of that found for $\chi$~Per (see Figs.\ \ref{fig:PMpi} and \ref{fig:Ppi}), left panels), especially when one considers the zero point parallax error of \hbox{$\pi_{\rm ZP}=-0.05\pm0.03$}.  

The Gaia DR2 distance estimate for $\chi$~Per, \dchiper, is consistent with the previous studies described above. The internal dispersion on the average cluster parallax is extremely precise ($\pm1.3\%$), and so the uncertainty on the absolute distance is dominated by that on $\pi_{\rm ZP}$. Even so, the absolute distance is precise to $\pm$7\%. Combined with the agreement with the two independent studies described above, we can consider the distance to $\chi$~Per to be extremely well constrained.

\subsection{NGC~7419}
In contrast to $\chi$~Per, the various distance estimates for NGC~7419 found in the literature span a broad range of values. Several studies exist which in one way or another fit the main sequence and/or spectroscopic parallaxes to the brightest main-sequence stars \citep{Beauchamp94,Caron03,Subramaniam06,Joshi08,Marco-Negueruela13}, but which find distances ranging from 1.7-4.0kpc. Further, despite having one extreme RSG as a cluster member (MY~Cep), which is a known maser emitter \citep[e.g.][]{Verheyen12}, there is no parallax measurement for this star, and so this object cannot be used to resolve the controversy. 

We find an average parallax of \pingc\ from the `clean' sample of OB stars. We note that this value is robust to the details of which stars in the original OB sample we include in the averaging. As seen in \fig{fig:PMpi}, most OB stars in the plane of the cluster have similar proper motions. Irrespective of how harsh we make the proper motion cuts, we always obtain the same average parallax within the errors. The parallax translates to a distance of \dngc, which is consistent with the mean of the measurements described in the previous paragraph. Again, the dominant source of error is that on $\pi_{\rm ZP}$.


\subsection{Westerlund 1}
As summarised recently by A19, there have been numerous and wide-ranging distance estimates for Wd~1. Of the contemporary measurements, whether they be based on the assumed intrinsic luminosities of B-supergiants \citep{Crowther06}, a kinematic distance based on the radial velocity of the HI gas \citep{Kothes-Dougherty07}, or fitting the (pre-) main sequence \citep{Brandner08}, all seem to converge on $\sim$4kpc. 

In the past year, other authors have looked at Wd~1's parallax information in Gaia. \citet{Clark18} quoted an average parallax of $\pi=$0.21-0.24 (assuming $\pi_{\rm ZP} = -0.05$), but commented that the errors on the parallaxes of individual stars meant that one could only say that the cluster was consistent with the recent estimates. A19 went further, and attempted to model the large number of stars in the plane of Wd~1 into field and cluster components based on the observed parallax distribution. For the cluster component, they found $\bar{\pi} = 0.31\pm0.04$mas, and a distance of $d=3.2\pm0.4$kpc. Though consistent with the canonical `4kpc' distance to Wd~1 to within the errors, these authors argued that this nearer distance would require an older age for the cluster, and would have profound implications for the origins of Wd~1's many post main-sequence objects. 

The methodology of our study is different enough to that of A19 to be complimentary. In A19, they assume that the `core-region' is dominated by cluster stars, which gives them a very large sample. This sample inevitably contains foreground contaminants, which these authors then attempt to model out. Though we have fewer stars on which to base $\bar{\pi}$, the spectroscopic and proper-motion selection function mean that we have a very high membership probabilities for all stars in our sample. This means that we do not have to fit for the spatial distribution of the field star population, and so have at least three fewer free parameters (the cluster and field star densities, and the lengthscale for the field star distribution function). We find an average parallax to Wd~1 of \piwdone, where we have applied the zero-point offset of $\pi_{\rm ZP} = -0.05$, but have {\it not} yet included the error on $\delta\pi_{\rm ZP}$ in the total uncertainty. This agrees to within $\sim2\sigma$ of that found by A19 and \citet{Clark18}, once the same value of $\pi_{\rm ZP}$ is adopted. There is a variation in our measurement of $\bar{\pi}$ of $\pm5\%$ depending on how tightly we perform the proper motion cleaning and whether we incorporate the excess astrometric noise into the parallax error, though this is well within the quoted $1\sigma$ uncertainty. The impact of this variation on the inferred distance is discussed next. 

The posterior distribution on distance $P_d$ is plotted in \fig{fig:dist}. Our result on the distance to Wd~1 is \dwdone. The variation of $\bar{\pi}$ caused by how aggressively we perform the proper motion cleaning can cause the inferred distance to vary between 3.6--4.1kpc. As with the average parallax, our distance estimate is within the errors of that of A19, but systematically higher, and with conspicuously larger errors despite the errors on $\bar{\pi}$ being comparable. We are unable to provide a definitive explanation for this, but we speculate that it is caused by our treatment of $\delta\pi_{\rm ZP}$. Here, we say that the error on Gaia's zero-point parallax offset affect all stars equally, since the angular scale for variations in $\pi_{\rm ZP}$ \citep[$\sim$1\degr, ][]{Lindegren18} is larger than the radius of Wd~1. This means that $\delta\pi_{\rm ZP}$ sets a hard limit on the precision of any measurement of absolute distance, regardless of the number of stars used to define the cluster average parallax. 

Our measurement of Wd~1's distance therefore places it close to the $\sim$4kpc found by previous studies. Furthermore, the uncertainty on this distance is roughly double that quoted by A19. We argue that this error bar cannot be reduced without a better characterization of Gaia's zero-point parallax offset. In addition, the chromatic calibration of Gaia in DR2 is still in its initial stages, and so this may be a further source of systematic error for heavily reddened clusters such as Wd~1.

\section{Summary} \label{sec:conc}
Using Gaia Data Release 2, we have reappraised the distances to three Milky Way young massive star clusters using the average parallaxes of their hot star cluster members. For $\chi$~Per, we find a distance in excellent agreement with earlier estimates (\dchiper). For NGC~6419, our distance is right in the middle of the varied estimates present in the literature (\dngc). Finally, for Westerlund~1, our distance of \dwdone\ is consistent with previous estimates, though with a larger error than a recent paper which also uses Gaia DR2 parallaxes. We argue that our errors are the more realistic given the current uncertainties on Gaia's zero-point parallax offset. This implications for these revised distances on the cluster ages are discussed in a companion paper \citep{Beasor2019}.

\section*{Acknowledgements}
We thank Mojgan Aghakhanloo, Simon Clark, Paul Crowther, Sebastian Kamann, Jeremiah Murphy, Ignacio Negueruela and the anonymous referee for useful discussions and constructive comments.  

This work has made use of data from the European Space Agency (ESA) mission
{\it Gaia} (\url{https://www.cosmos.esa.int/gaia}), processed by the {\it Gaia}
Data Processing and Analysis Consortium (DPAC,
\url{https://www.cosmos.esa.int/web/gaia/dpac/consortium}). Funding for the DPAC
has been provided by national institutions, in particular the institutions
participating in the {\it Gaia} Multilateral Agreement.




\bibliographystyle{mnras}
\bibliography{/Users/astbdavi/Google_Drive/drafts/biblio} 


\bsp	
\label{lastpage}
\end{document}